\documentclass{article}
\usepackage{arxiv}
\usepackage[utf8]{inputenc} 
\usepackage[T1]{fontenc}    
\usepackage{hyperref}       
\usepackage{url}            
\usepackage{booktabs}       
\usepackage{amsfonts}       
\usepackage{nicefrac}       
\usepackage{lipsum}
\usepackage{fancyhdr}       
\usepackage{graphicx}       
\graphicspath{{media/}}     
\usepackage{xcolor}
\usepackage[utf8]{inputenc} 
\usepackage[T1]{fontenc}  
\usepackage{ragged2e}
\usepackage{amssymb}
\usepackage{amsthm}
\usepackage{amsmath,mathrsfs}
\usepackage{color}
\usepackage{graphicx}
\usepackage{comment}
\usepackage{mathtools}
\usepackage{lineno}
\usepackage{algorithm}
\usepackage{algorithmicx}
\usepackage{algpseudocode}
\usepackage{subcaption}
\usepackage{graphicx}
\usepackage[numbers]{natbib}
\usepackage{cleveref}
\usepackage{pdflscape}
\usepackage[normalem]{ulem}
\usepackage{float}

\usepackage{setspace}
\title{\textbf{Digital twin with automatic disturbance detection for an expert-controlled SAG mill}
}

\author{
  Paulina Quintanilla \\
  Department of Chemical Engineering\\
  Brunel University of London \\
  London, United Kingdom\\
  \texttt{paulina.quintanilla@brunel.ac.uk} \\
     \And
  Francisco Fern\'andez \\
  Departamento de Ingenier\'ia Qu\'imica y Ambiental\\
  Universidad T\'ecnica Federico Santa Mar\'ia \\
  Santiago, Chile\\
     \And
  Cristobal Mancilla, Mat\'ias Rojas \\
  Departamento de Ingenier\'ia Qu\'imica y Ambiental\\
  Universidad T\'ecnica Federico Santa Mar\'ia \\
  Advanced Systems, SGS Minerals \\
  Santiago, Chile\\
     \And
  Daniel Navia \\
  Departamento de Ingenier\'ia Qu\'imica y Ambiental\\
  Universidad T\'ecnica Federico Santa Mar\'ia \\
  Santiago, Chile\\
  \texttt{daniel.navia@usm.cl} \\
}

\begin{document}
\maketitle

\begin{abstract}
This study presents the development and validation of a digital twin for a semi-autogenous grinding (SAG) mill controlled by an expert system. The digital twin integrates three key components of the closed-loop operation: (1) fuzzy logic for expert control, (2) a state-space model for regulatory control, and (3) a recurrent neural network to simulate the SAG mill process. The digital twin is combined with a statistical framework for automatically detecting process disturbances (or critical operations), which triggers model retraining only when deviations from expected behaviour are identified, ensuring continuous updates with new data to enhance the SAG supervision. The model was trained with 68 hours of operational industrial data and validated with an additional 8 hours, allowing it to predict mill behaviour within a 2.5-minute horizon at 30-second intervals with errors smaller than 5\%. 
\end{abstract}

\section{Introduction}
\label{sec: introduction}

Rapid advancements in digital technologies have significantly transformed various industrial processes, introducing innovative methods for control and monitoring. One notable advancement is the development of digital twins, which have gained substantial traction in recent years. A digital twin is a key technology of Industry 4.0 that integrates a physical system with its virtual counterpart through models, sensors, and data \cite{MELESSE2020267}. The digital twin continuously adapts to operational changes, forecasting future performance using real-time data that can be used for process monitoring and control.

Despite increasing interest, the application of digital twins in Semi-Autogenous Grinding (SAG) mill operations remains underexplored \cite{GHASEMI2024108733}. SAG mills are crucial in the comminution process, grinding large rocks into smaller particles for subsequent processing stages. The efficiency and stability of SAG mills directly impact the overall performance of mineral processing plants. Given the high operational costs and energy consumption associated with SAG mills, optimal performance is crucial to the economic viability of mining operations. Traditionally, expert control systems have been used to control and stabilize SAG mill operations by leveraging predefined rules and real-time data. Previous research has developed dynamic simulations and control frameworks for SAG mills, including real-time control systems to stabilize mill operations \cite{GUERRERO201661}, and fault diagnosis approaches using process data \cite{WAKEFIELD2018132}. Currently, to the best of the authors' knowledge, there is no existing framework in the literature that integrates the expert system, regulatory control, and a dynamic model of the process into a more robust and proactive control environment.

This paper addresses a key gap in the current literature by developing a digital twin that integrates expert control systems, regulatory control, and a model of the SAG mill process. The digital twin is based on a statistical framework that predicts critical variables such as bearing pressure and motor power based on three inputs: tonnage, solids percentage, and mill speed. The robustness of the model and automatic detection algorithm is assessed by simulating multiplicative disturbances in the controlled variables.

\section{Methodology}
\label{sec: methodology}

\subsection{Model components}
\begin{landscape}
\begin{figure}
    \includegraphics[scale=0.9]{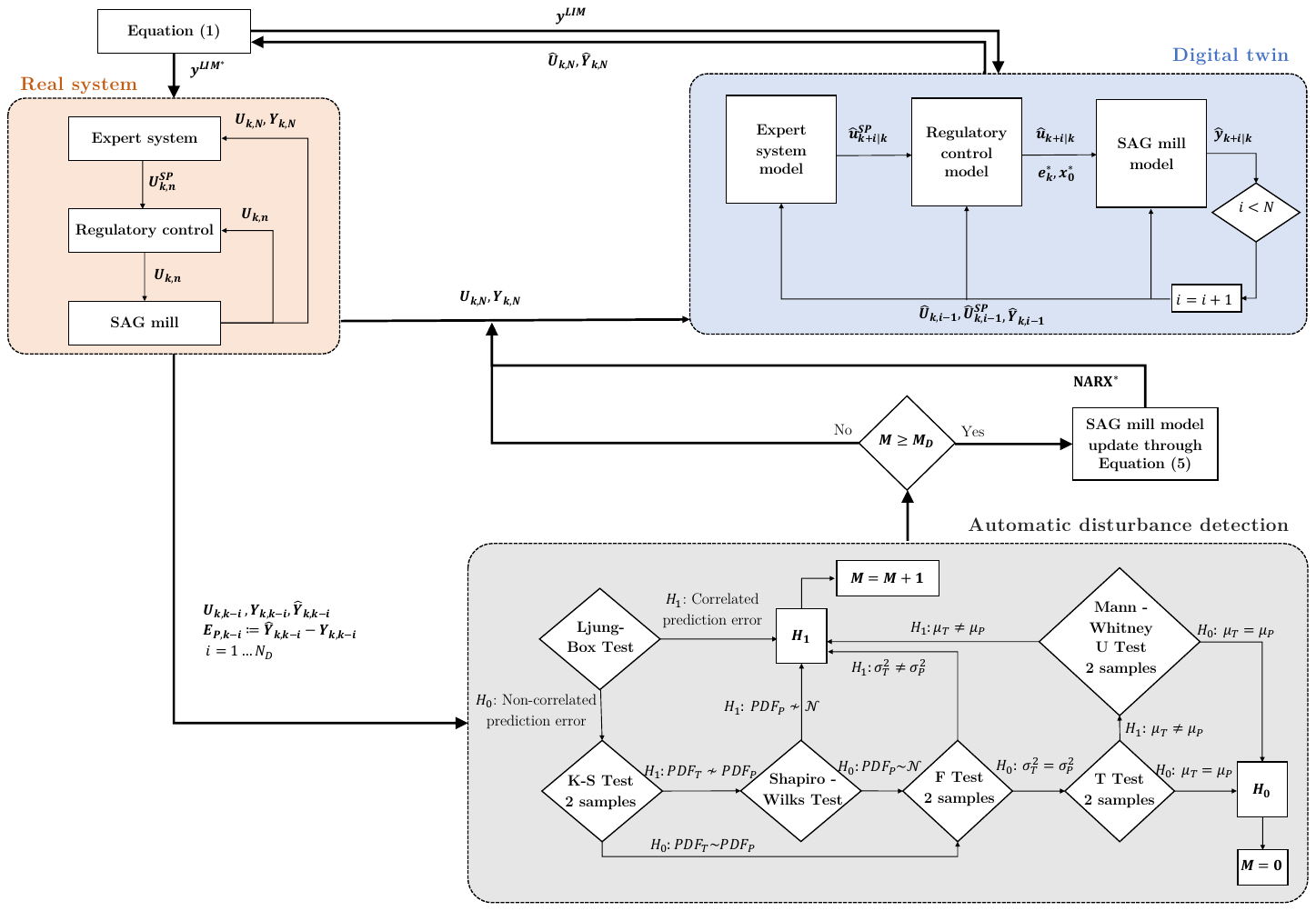}
    \caption{Proposed supervisory system architecture, featuring a digital twin and an automatic disturbance detection module. The 'real system' illustrates the closed-loop operation of the actual SAG mill, while the 'digital twin' simulates this behavior in real-time.  Subscripts: P denotes prediction error, and T denotes training error.}
    \label{fig: system}
\end{figure}
\end{landscape}

The studied system is a closed-loop SAG system (denoted as \textit{real process} in \Cref{fig: system}), which comprises three parts: 
\begin{enumerate}
\item \textbf{Expert control system}: It relies on operational conditions determined by the operational limits of the controlled variables (CV): bearing pressure ($y_1$) and motor power ($y_2$) of the mill, which can be adjusted externally. It functions as an upper hierarchical layer with an algorithm that employs the current and past values of controlled and manipulated variables (MV) to recognize the current mode of SAG operation and suggest adjustments to the MV setpoints ($u^{SP}$).

\item \textbf{Regulatory control}: This lower hierarchical layer is responsible for implementing changes proposed by the expert control system by modifying the MV values: tonnage ($u_1$), solids percentage ($u_2$), and mill speed ($u_3$) to reach their setpoint values.

\item \textbf{SAG mill:} It is the actual process to be optimized, which was simulated using recurrent neural networks as described in \Cref{subsec: sag-model}. 
\end{enumerate}

A new variable, $\boldsymbol{y}^{LIM}$, is defined to represent the degrees of freedom available in the closed-loop real system. This implies that it is possible to implement a supervisory system over the expert control system to determine the value of $\boldsymbol{y}^{LIM}$ to optimize the expected dynamic response by solving the problem in  \Cref{eq: ylim-opt} and implementing the results using a moving horizon strategy. Note that defining $\boldsymbol{y}^{LIM}$ as a variable that can be adjusted during operation provides an additional layer of flexibility, allowing the expert system to operate more aggressively under less critical conditions, or more conservatively as critical limits are approached. Thus, the digital twin simulates the dynamic response of the SAG mill to adjustments in $\boldsymbol{y}^{LIM}$, providing valuable insight into how the system can be optimized under various operational scenarios.

\newpage

\newpage

\begin{equation}
\label{eq: ylim-opt}
\begin{aligned}
& \min _{\boldsymbol{y}^{L I M} \in \mathcal{Y}^{L I M}} f\left(\widehat{\boldsymbol{Y}}_{k, N}, \widehat{\boldsymbol{U}}_{k, N}, \widehat{\boldsymbol{U}}_{k, N}^{S P}\right) \\
& \text { s.t.: } \\
& \widehat{\boldsymbol{y}}_{k \mid k}=\boldsymbol{h}\left(\boldsymbol{Y}_{k, m}, \widehat{\boldsymbol{U}}_{k, 0}, \boldsymbol{U}_{k, n}, \widehat{\boldsymbol{U}}_{k, 0}^{SP}\right) \\
& \widehat{\boldsymbol{y}}_{k+i \mid k}=\boldsymbol{h}\left(\widehat{\boldsymbol{Y}}_{k, i-1}, \boldsymbol{Y}_{k,m-i}, \widehat{\boldsymbol{U}}_{k, i}, \boldsymbol{U}_{k, n-i}, \widehat{\boldsymbol{U}}_{k, i}^{S P}\right), \quad i=1 \ldots N \\
& \widehat{\boldsymbol{Y}}_{k, i}:=\left[\widehat{\boldsymbol{y}}_{k \mid k}, \ldots, \widehat{\boldsymbol{y}}_{k+i \mid k}\right], \widehat{\boldsymbol{U}}_{k, i}:=\left[\widehat{\boldsymbol{u}}_{k \mid k}, \ldots, \widehat{\boldsymbol{u}}_{k+i \mid k}\right], \widehat{\boldsymbol{U}}_{k, i}^{S P}:=\left[\widehat{\boldsymbol{u}}_{k \mid k}^{S P}, \ldots, \widehat{\boldsymbol{u}}_{k+i \mid k}^{S P}\right], \quad i=0 \ldots N \\
& \boldsymbol{Y}_{k, j}:=\left[\boldsymbol{y}_{k-1}, \ldots, \boldsymbol{y}_{k-j}\right], \quad j=1, \ldots, m, \quad \boldsymbol{U}_{k, j}:=\left[\boldsymbol{u}_{k-1}, \ldots, \boldsymbol{u}_{k-j}\right], \quad j=1, \ldots, n \\
& \widehat{\boldsymbol{y}}_{k+i \mid k} \in \mathcal{Y}, \widehat{\boldsymbol{u}}_{k+i \mid k} \in \mathcal{U}, \quad i=0 \ldots N 
\end{aligned}
\end{equation}
\\
\noindent where $k$ represents the current instant, $\mathbf{y}_{k-i}$ and $\textbf{u}_{k-i}$ correspond to the measured CV and MV at $i$ previous instants relative to $k$, and $\boldsymbol{\hat{y}}_{k+i|k}$, $\boldsymbol{\hat{u}}_{k+i|k}$, and $\boldsymbol{\hat{u}}_{k+i|k}^{SP}$ denote the corresponding variables, predicted $i$ instants into the future relative to $k$. The digital twin of the process is represented by \textit{$\textbf{h}$}, which can be understood as a discrete model that allows predicting the behavior of the system's variables in the closed-loop of \Cref{fig: system}, while $f$ is the objective function of the process to optimize. In \Cref{eq: ylim-opt}, $\mathbf{\mathcal{Y}}$, and $\mathbf{\mathcal{U}}$ describe the feasible region of the optimization problem, while $n$ and $m$ correspond to the previous instants required by the dynamic model to simulate the next value. $N$ corresponds to the prediction horizon.

The implementation of the supervisory system requires adaptability or learning capability. Learning consists of the systematic update of the process model to avoid degradation in its predictive capacity, associated with the inherent mismatch related to the assumptions made to obtain said model \cite{darby2011rto}. The learning process can be implemented by retraining the model so that it adequately predicts a window of recent data. However, overfitting should be avoided as it could lead to learning behavior associated with measurement noise. 
A digital twin was developed to emulate the architecture of the process and was implemented in Python 3.7. It is composed of three modules connected in series with models of each of the sections of the closed-loop system (see "Digital twin" module in \Cref{fig: system}).

\subsubsection{Expert control system model} 

The expert control system model uses a fuzzy logic approach, taking the values and slopes of the CVs and MVs and transforming them into fuzzy sets using triangular membership functions. The operating state of the process is determined by the maximum value of each fuzzy set, which is associated with a criticality hierarchy. The expert system recommends a hierarchy of actions, with tonnage reduction as the last step. Tonnage is reduced when the SAG is operating at or near critical levels, detected by high pressure and/or power values. Tonnage reduction has a significant impact on reducing SAG pressure and power but decreases the mill's processing capacity. Based on this hierarchy, the necessary actions are determined and then transformed into changes in the manipulated variables using Sugeno's approach \cite{sugeno1985industrial}. 

\subsubsection{Regulatory control model} 

It is important to note that the expert system uses the current values of the MVs and their dynamic behavior to make adjustments. The dynamics of these variables do not always align with their setpoints, as demonstrated through previous data analysis. The state-space model captures this dynamic response and is necessary as input to the expert system model. The regulatory control model simulates the dynamic response of control loops of MVs using a state-space model as shown in \Cref{eq: eq2}. The state-space model is a discrete version of a linear system of algebraic-differential equations that describes the dynamic dependence of process inputs ($\boldsymbol{\hat{u}}^{S P}$) with outputs ($\boldsymbol{\hat{u}}$), through state variables ($\boldsymbol{\hat{x}}$) that determine the future behaviour of a given system, when present state of the system and inputs are known \cite{dorf2022modern}, 

\begin{equation}
\begin{aligned}
& \boldsymbol{\hat{x}}_{k+i+1 \mid k}=\mathbf{A} \boldsymbol{\hat{x}}_{k+i \mid k}+\mathbf{B} \boldsymbol{\hat{u}}_{k+i \mid k}^{S P}+\mathbf{K} \boldsymbol{e}_k, \quad \boldsymbol{\hat{x}}_{k \mid k}=\boldsymbol{\hat{x}}_0 \\
& \boldsymbol{\hat{u}}_{k+i+1 \mid k}=\mathbf{C} \boldsymbol{\hat{x}}_{k+i \mid k}+\mathbf{B} \boldsymbol{\hat{u}}_{k+i \mid k}^{S P}+\boldsymbol{e}_k,
\label{eq: eq2}
\end{aligned}
\end{equation}

\noindent where $\boldsymbol{\hat{x}}_{k+i|k}$ corresponds to the vector of the state variables of the model, predicted $i$ instants into the future relative to $k$. Matrices $\mathbf{A}$, $\mathbf{B}$, $\mathbf{C}$, and $\mathbf{K}$ contain the model parameters, which are obtained by solving \Cref{eq:4}, $\boldsymbol{e}_k$ represents a constant additive disturbance vector, and $\boldsymbol{\hat{x}}_0$ is the initial value of the states.

\subsubsection{SAG mill model} 
\label{subsec: sag-model}
The SAG mill model predicts the dynamic response of the controlled variables to the changes simulated by the regulatory control model. To account for the effects of the system's free and forced response, a nonlinear autoregressive exogenous (NARX) model based on neural networks \cite{hopfield1982neural} was used, incorporating dependencies on the past values of the MVs and CVs. The dependency of the implemented NARX model is summarized in \Cref{eq:3}, where \textbf{\textit{g}} represents the model composed of the interconnection of the recurrent neural network layers.

\begin{equation}
\label{eq:3}
\widehat{\boldsymbol{y}}_{\boldsymbol{k}+\boldsymbol{i} \mid \boldsymbol{k}}=\boldsymbol{g}\left(\widehat{\boldsymbol{Y}}_{\boldsymbol{k}, \boldsymbol{i}-1}, \boldsymbol{Y}_{\boldsymbol{k}, \boldsymbol{m}-\boldsymbol{i}}, \widehat{\boldsymbol{U}}_{\boldsymbol{k}, \boldsymbol{i}}, \boldsymbol{U}_{\boldsymbol{k}, \boldsymbol{n - i} \boldsymbol{i}}\right), \quad i=0 \ldots N
\end{equation}

\subsection{Model identification and training}
The data used for parameter estimation of the models in \Cref{eq: eq2} and \Cref{eq:3} were extracted from 6 months of historical operation of the SAG, obtained every 5 seconds. This database was filtered using MySQL according to the following criteria: SAG in operation, feed above the operational minimum, percentage of solids above the operational minimum, and expert control system online. The resulting data sets were filtered using a moving median between 6 contiguous data points, obtaining data sets with a sampling time of 30 seconds. The training data were selected from the largest resulting set, with 8156 time instants, equivalent to 68 hours of operation. The next largest data set (1013 time instants, equivalent to 8 hours of operation) was defined as the test data set and was used for the validation of the trained digital twin.

\subsubsection{Regulatory control identification:} The identification of the regulatory control model is summarized in \Cref{eq:4}, where the sum of the objective function and the model applies to the entire training data horizon ($N_H$).

\begin{equation}
\label{eq:4}
\begin{aligned}
& \min _{ \boldsymbol{A},  \boldsymbol{B},  \boldsymbol{C},  \boldsymbol{K}, \boldsymbol{\widehat{x}}_0} \sum_{k=1 \ldots N_H}\left\|\boldsymbol{u}_{k}-\widehat{\boldsymbol{u}}_{k \mid k}\right\|_2 \\
& \text { s.t.: Eq. (2), with } \boldsymbol{e}_{k} =\mathbf{0}
\end{aligned}
\end{equation}

To solve \Cref{eq:4}, it is necessary to define the model order, which represents the number of previous time instants required for prediction. The procedure to determine this value involves first setting the model order and then estimating the parameters. This process is repeated for different model orders, and the smallest order that provides a fit comparable to higher orders, without significant improvement, is selected.

\subsubsection{SAG mill model identification}

The identification of the SAG mill model is summarized in \Cref{eq:5}, where the sum of the objective function and the model applies to the entire training data horizon. NARX summarizes the parameters of the neurons: input weights, bias, and activation. Only one hidden layer was used.

\begin{equation}
\label{eq:5}
\begin{aligned}
& \min _{\text {NARX }} \sum_{k=1 \ldots N_H}\left\|\boldsymbol{y}_{k}-\widehat{\boldsymbol{y}}_{k \mid k}\right\|_2 \\
& \text { s.t. : Eq. (3) }
\end{aligned}
\end{equation}

To solve \Cref{eq:5}, the number of neurons in the hidden layer and previous instants of the MVs and CVs must be defined. The following procedure was implemented for their estimation: 1) set the number of previous instants and neurons in the hidden layer, 2) estimate parameters. This procedure is performed for different combinations of previous instants and the number of neurons, and the combination with the smallest number of previous instants and hidden neurons, whose fit does not significantly improve compared to models with larger combinations, is selected.

\subsection{Automatic disturbance detection}
The automatic disturbance detection acts as a learning module to update the models of the digital twin for regulatory control and the SAG mill (see "Automatic disturbance detection" module in \Cref{fig: system}). The update of the expert control system model is not considered, as it is subject to changes in the real expert control system, a condition controlled externally. The learning of the regulatory control model consists of estimating $\boldsymbol{\hat{x}}_0$ and $\mathbf{e}_k$ by solving \Cref{eq:6} each time the digital twin is used (assumed at each instant $k$). The estimation horizon ($N_E$) corresponds to the number of previous instants used. In \Cref{eq:6}, $ \boldsymbol{A}^*$, $ \boldsymbol{B}^*$, $ \boldsymbol{C}^*$, and $ \boldsymbol{K}^*$ represent the results of \Cref{eq:4}.

\begin{equation}
\label{eq:6}
\begin{aligned}
& \min _{\boldsymbol{\widehat{x}}_0, \mathbf{e}_k} \sum_{i=1 \ldots N_E}\left\|\boldsymbol{u}_{\boldsymbol{k}-\boldsymbol{i}}-\widehat{\boldsymbol{u}}_{\boldsymbol{k}-\boldsymbol{i} \mid \boldsymbol{k}}\right\|_2 \\
& \text { s.t. : Eq. (2) with } \boldsymbol{A}^*, \boldsymbol{B}^*, \boldsymbol{C}^*, \boldsymbol{K}^*
\end{aligned}
\end{equation}

The learning of the SAG mill model consists of detection and retraining to avoid overfitting with measurement noise since neural networks are universal regressors. The automatic disturbance detection module evaluates whether the model prediction deteriorates significantly compared to the training results at each instant $k$. For this, the characteristics of the prediction error between two data sets are compared: training (T) and the one obtained in a window of $N_D$ recent data points (P). This comparison is done through two-sided hypothesis tests for error correlation, probability distribution function (PDF), variance, and mean. If no null hypothesis is rejected, the counter $M=0$ is set; otherwise, the counter is updated $M=M+1$. This procedure is performed at each sampling instant. If the value of $M$ exceeds the detection threshold $M_D$, then the SAG mill model is retrained by solving \Cref{eq:5} with the most recent $N_E$ process data that meets the training criteria. This iterative process to find the value for $M_D$ is described in \Cref{alg:alg}.

\begin{algorithm}
\caption{Iterative process to find $M_D$ (threshold for retraining)}
\label{alg:alg}
\begin{algorithmic}[1]
\State Initialize: 
\State $k = 0$ \Comment{Iteration counter}
\State $M_D^k = N_D + 1$ \Comment{Initial threshold based on data window size $N_D$}
\While{True}
    \State \textbf{Step 1:} Simulate the digital twin using the test dataset 
    \State \textbf{Step 2:} Obtain the moving data window of prediction errors
    \State Simulate digital twin with a test dataset
    
    \If{$M \geq M_D^k$} 
        \State \textbf{Step 3:} Increment the threshold and iteration counter
        \State $M_D^{k+1} = M_D^k + 1$
        \State $k = k + 1$
    \Else
        \State \textbf{Step 4:} Detected threshold $M_D^k$
        \State \textbf{Break}
    \EndIf
\EndWhile

\State \textbf{Output:}
\State Detected threshold value: $M_D^k$

\State \textbf{Define retraining horizon:}
\State $N_H = \text{Maximum available dataset for initial training}$
\State $N_E = M_D + N_E^+$ \Comment{Where $N_E^+ \geq 0$ represents additional data before retraining}

\end{algorithmic}
\end{algorithm}

\newpage

\section{Results and discussions}
\label{sec: results}

\subsection{Digital twin simulation}
\begin{figure}[htp!]
    \centering
    \includegraphics[width=1\textwidth]{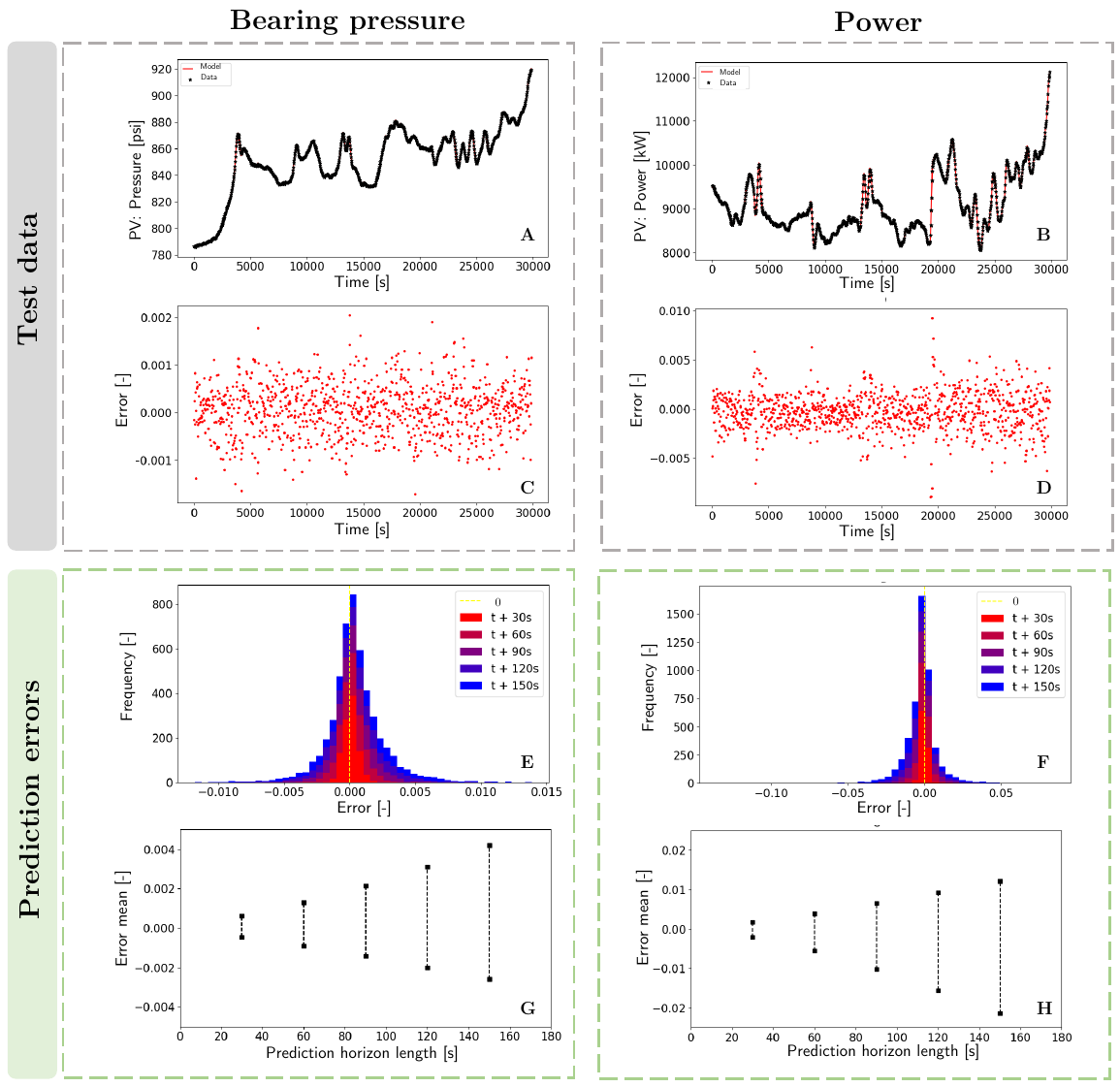}
    \caption{Digital twin simulation results: (A-B) test data and predictions; (C-D) prediction errors; (E-F) histogram of prediction errors; (G-H) prediction error intervals.}
    \label{fig: fig2}
\end{figure}

\Cref{fig: fig2} shows the output of the NARX model simulating the process in a closed-loop configuration. This means that at each discrete time step, the predictions of pressure and power are obtained by sequentially solving the three integrated modules (expert system, regulatory control model, and SAG mill model), using the current value of $\boldsymbol{y}^{LIM}$  and the previously measured values required by the SAG mill model. The SAG mill process was simulated using NARX models with 1 hidden layer composed of 2 neurons, and 12 previous time instances were used. \Cref{fig: fig2} A-B represents the test data and their predictions, \Cref{fig: fig2} C-D presents the prediction errors, while \Cref{fig: fig2} E-H summarizes the main characteristics of the digital twin's prediction error for 5 prediction instants. Regarding error distribution, the histograms in \Cref{fig: fig2} E-F indicate a normal pattern centred around zero at all prediction times. As for the prediction errors, \Cref{fig: fig2} G-H demonstrates that the bearing pressure and motor power predictions are also centered around zero. The dispersion increases as the prediction time frame extends due to the feedback of the simulated CV errors associated with the autoregressive nature of the NARX model. The prediction error for 2.5 minutes of prediction is expected to be within the interval \([-1\%, 1\%]\) for the bearing pressure and \([-5\%, 5\%]\) for power, with 99\% certainty. Considering that the data showed no correlation with the explanatory variables, it can be indicated that the digital twin adequately describes the real closed-loop system shown in \Cref{fig: system}.

\subsection{Automatic detection algorithm for learning}

The automatic detection algorithm employed a 30-sample analysis window to monitor changes in the correlation of key process variables and determine the need for model retraining. The threshold value, $M_D$, was set at 103 for bearing pressure and 181 for motor power, following the iterative process described in \Cref{alg:alg}. The algorithm was tested using synthetic perturbations to simulate expected operational changes: gradual wear of the mill’s metal liner over one and five months, and an increase in ore hardness by 10\%. These modifications assumed a 2\% monthly liner wear, inversely proportional to bearing pressure and directly proportional to the mill’s rotation speed. At the same time, the ore hardness increase was directly proportional to both bearing pressure and motor power. These simulated scenarios were designed to validate the algorithm’s ability to detect perturbations affecting the data correlation.

\begin{figure} [H]
    \centering
    \includegraphics[width=1\textwidth]{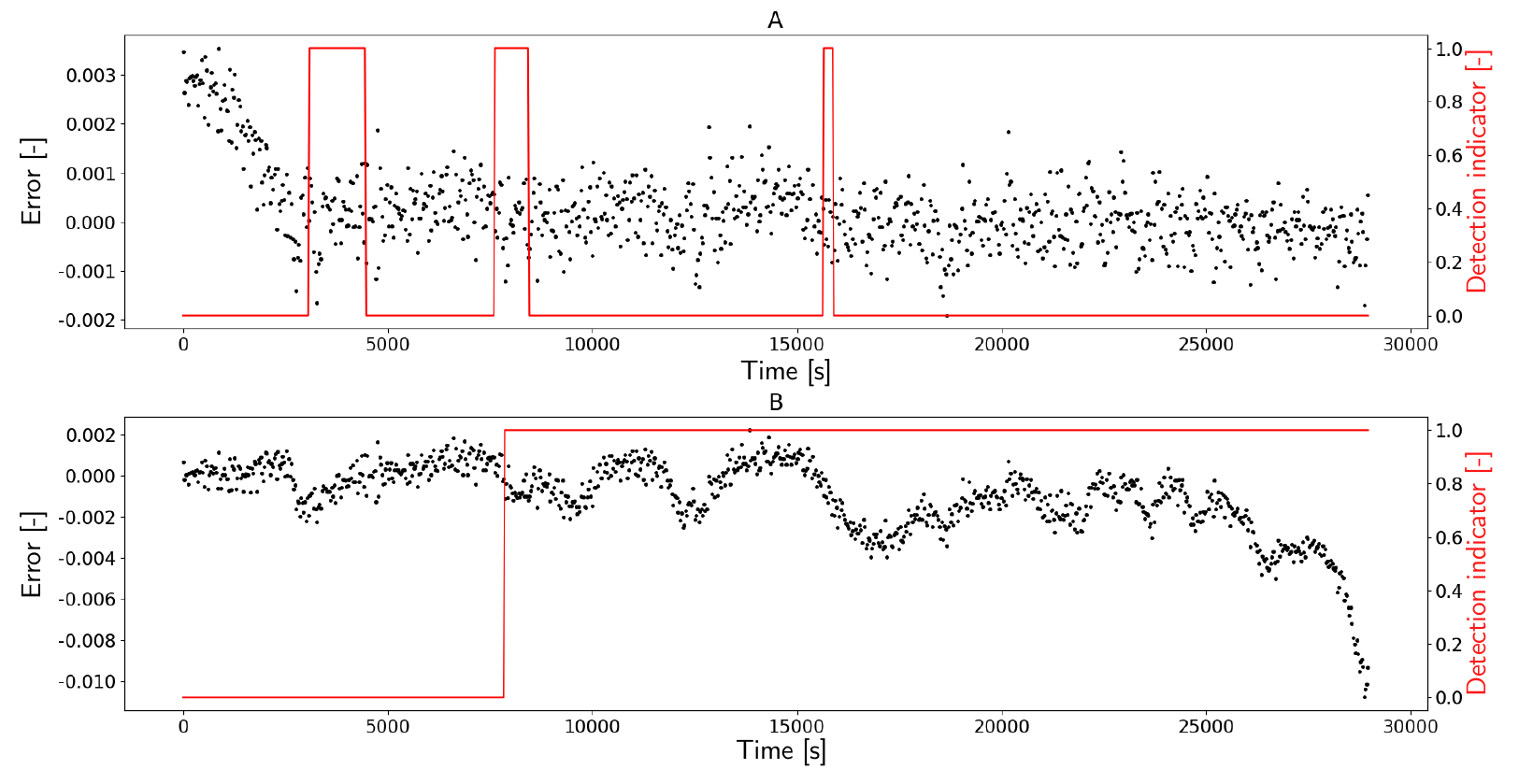}
\caption{Proportional errors of bearing pressure prediction and automatic learning detection indicator for two disturbance scenarios: (A) mill liner wear after 5 months of operation, (B) 10\% increase in ore hardness.}
    \label{fig: fig3}
\end{figure}

In the scenarios tested, the algorithm did not indicate the need for retraining the motor power model, and no updates were triggered during the 1-month liner wear scenario. However, the detection system successfully identified the need for retraining for the 5-month liner wear and ore hardness increase scenarios, as shown in \Cref{fig: fig3}. Specifically, in the 5-month wear scenario, the detection system flagged three instances for retraining due to consistent underestimation of pressure, which indicated a mismatch between the model predictions and actual bearing pressure values (\Cref{fig: fig3}A). This was attributed to the cumulative 10\% increase in pressure caused by the disturbance. In the ore hardness scenario (\Cref{fig: fig3}B), the detector was activated after 7000 seconds due to the loss of correlation between prediction errors and actual measurements, reflecting the model's inability to capture the increased demand on motor power caused by the harder ore.

\section{Conclusions}
\label{sec: conclusions}
This study presents a comprehensive digital twin framework for closed-loop SAG mill optimization, integrating expert control, regulatory control, and a recurrent neural network for process simulation. The proposed digital twin is designed to evaluate and adjust the upper limits of control actions by increasing or decreasing the aggressiveness of proposed control strategies. This is aimed at reducing the potential impact of critical conditions without needing to replace the supervisory control system. 

The digital twin successfully predicts the SAG mill's behavior under various disturbances, achieving a maximum prediction error of 5\% for a 2.5-minute horizon and less than 1\% for a 30-second prediction. The automatic disturbance detection system is designed to prevent overfitting, ensuring that the supervisory system maintains its performance without significant degradation due to expected process changes. Upon detecting disturbances, the retraining procedure effectively adjusts the model, ensuring that it continues to accurately reflect the current process state and remains reliable and adaptable. Future work will focus on incorporating this digital twin into real-time optimization strategies. This approach provides a solid foundation for advancing control systems in SAG mill operations.

 \bibliographystyle{unsrt} 
 \bibliography{cas-refs}

\begin{thebibliography}{1}

\bibitem{MELESSE2020267}
Tsega~Y. Melesse, Valentina~Di Pasquale, and Stefano Riemma.
\newblock Digital twin models in industrial operations: A systematic literature review.
\newblock {\em Procedia Manufacturing}, 42:267--272, 2020.
\newblock International Conference on Industry 4.0 and Smart Manufacturing (ISM 2019).

\bibitem{GHASEMI2024108733}
Zahra Ghasemi, Mehdi Neshat, Chris Aldrich, John Karageorgos, Max Zanin, Frank Neumann, and Lei Chen.
\newblock An integrated intelligent framework for maximising sag mill throughput: Incorporating expert knowledge, machine learning and evolutionary algorithms for parameter optimisation.
\newblock {\em Minerals Engineering}, 212:108733, 2024.

\bibitem{GUERRERO201661}
F.~Guerrero, J.~Bouchard, E.~Poulin, and D.~Sbarbaro.
\newblock Real-time simulation and control of a sag mill.
\newblock {\em IFAC-PapersOnLine}, 49(20):61--66, 2016.
\newblock 17th IFAC Symposium on Control, Optimization and Automation in Mining, Mineral and Metal Processing MMM 2016.

\bibitem{WAKEFIELD2018132}
Monitoring of a simulated milling circuit: Fault diagnosis and economic impact.
\newblock {\em Minerals Engineering}, 120:132--151, 2018.

\bibitem{darby2011rto}
Mark~L. Darby, Michael Nikolaou, James Jones, and Doug Nicholson.
\newblock Rto: An overview and assessment of current practice.
\newblock {\em Journal of Process Control}, 21(6):874--884, 2011.

\bibitem{sugeno1985industrial}
M.~Sugeno.
\newblock {\em Industrial applications of fuzzy control}.
\newblock North-Holland; Sole distributors for the U.S.A. and Canada, Elsevier Science Pub. Co., Amsterdam; New York; New York, N.Y., U.S.A., 1985.

\bibitem{dorf2022modern}
R.C. Dorf and R.H. Bishop.
\newblock {\em Modern Control Systems}.
\newblock Pearson Education, Incorporated, 2022.

\bibitem{hopfield1982neural}
J.J. Hopfield.
\newblock Neural networks and physical systems with emergent collective computational abilities.
\newblock {\em Proceedings of the National Academy of Sciences of the United States of America}, 79:2554--2558, 1982.

\end{thebibliography}

\end{document}